\documentclass[sigconf]{acmart}

\AtBeginDocument{%
  \providecommand\BibTeX{{%
    \normalfont B\kern-0.5em{\scshape i\kern-0.25em b}\kern-0.8em\TeX}}}

\setcopyright{none}

\acmISBN{}
\acmPrice{}
\acmDOI{}
\acmYear{}

\acmConference[arXiv]{arXiv.org}{2001.06047v3}{25 Jan 2020}

\usepackage[acronym]{glossaries}

\newacronym{is}{IS}{Information System}
\begin{document}

\title[Expecting the Unexpected]{Expecting the Unexpected: Developing Autonomous-System Design Principles for Reacting to Unpredicted Events and Conditions}

\author{Assaf Marron}
\email{assaf.marron@weizmann.ac.il}
\orcid{0000-0001-5904-5105}
\affiliation{
  \institution{Weizmann Institute of Science}
  \streetaddress{234 Herzl St.}
  \city{Rehovot}
  \state{Israel}
  \postcode{76100}
}
\author{Lior Limonad}
\email{LIORLI@il.ibm.com}
\affiliation{
  \institution{IBM Research}
  \streetaddress{Mount Carmel}
  \city{Haifa}
  \state{Israel}
  \postcode{31905}
}
\author{Sarah Pollack}
\email{sarah.pollack@weizmann.ac.il}
\orcid{0000-0001-5904-5105}
\affiliation{%
  \institution{The Davidson Institute of Science Education \\
  Weizmann Institute of Science}
  \streetaddress{234 Herzl St.}
  \city{Rehovot}
  \state{Israel}
  \postcode{76100}
}
\author{David Harel}
\email{david.harel@weizmann.ac.il}
\affiliation{
  \institution{Weizmann Institute of Science}
  \streetaddress{234 Herzl St.}
  \city{Rehovot}
  \state{Israel}
  \postcode{76100}
}

\renewcommand{\shortauthors}{Marron, Limonad, Pollack, Harel}

\begin{abstract}
  When developing autonomous systems, engineers and other stakeholders make great efforts to prepare the system for all foreseeable events and conditions. However, such systems are still bound to encounter situations that were not considered at design time. For reasons like safety, cost, or ethics it is often highly desired that these new cases be handled correctly upon first encounter. In this paper, we first justify our position that there will always exist unpredicted events and conditions, driven  by, e.g., new inventions in  the real world, the diversity of world-wide system deployments and uses, and the possibility that multiple events that were overseen at design time will not only occur, but will occur together. We then argue that despite the unpredictability, handling such situations is indeed possible. Hence, we offer and exemplify design principles, which, when applied in advance, will enable systems to deal with unpredicted circumstances. We conclude with a discussion of how this work and a much-needed thorough study of the unexpected can contribute toward a foundation of engineering principles for developing trustworthy next-generation autonomous systems. 
\end{abstract}




\begin{teaserfigure}
  \includegraphics[width=\textwidth]{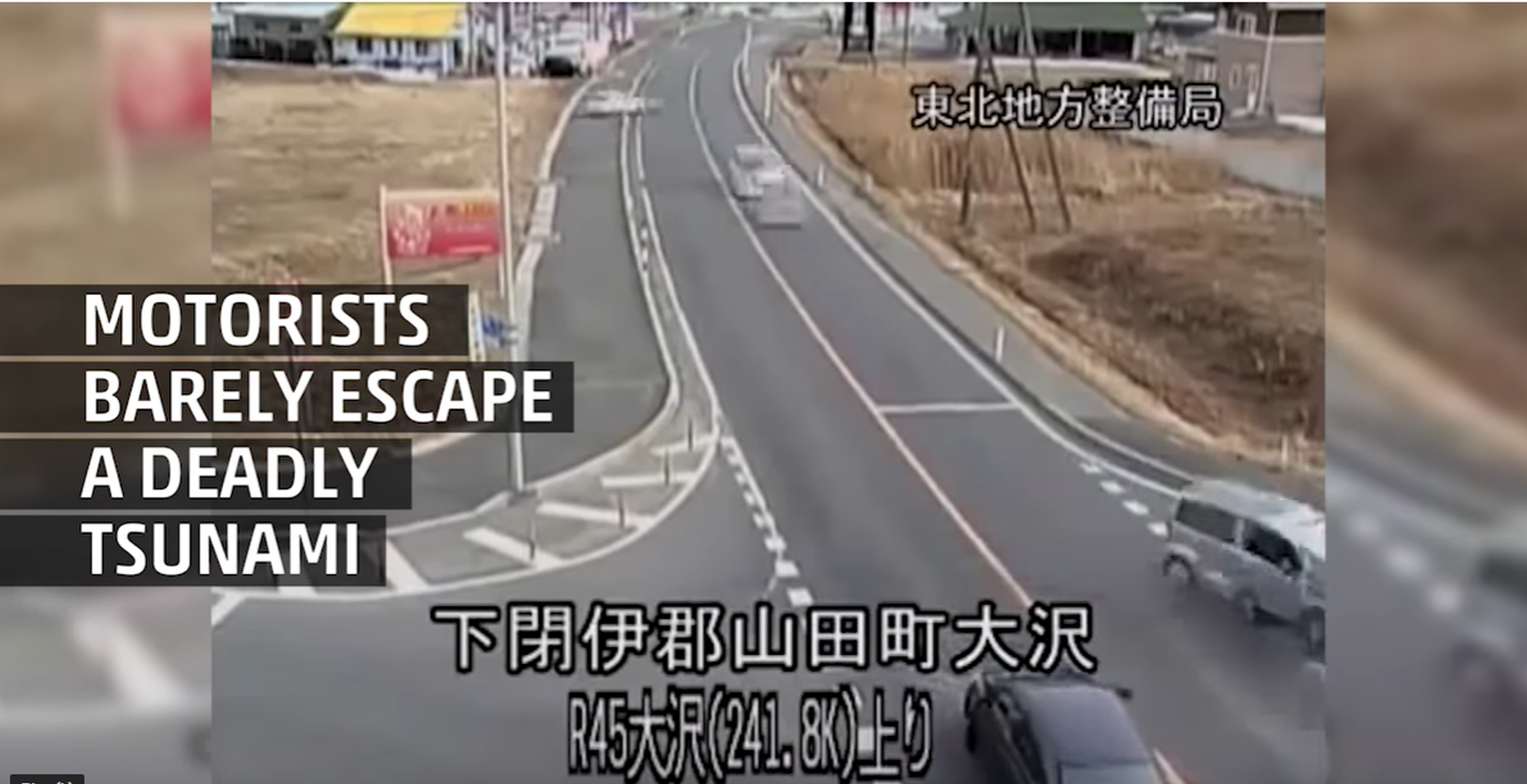}
  \caption{Japan, 2011. Motorists make U-turns and drive away 
  upon encountering a tsunami wave.\\ (A TV screen capture at \texttt{https://www.youtube.com/watch?v=coPW6unxeY8.)} 
}
\label{fig:tsunami}
\end{teaserfigure}

\maketitle

\section{Introduction}
Can the engineer of an autonomous system, such as a self-driving car or an in-hospital delivery-assistant robot, predict \emph{all} conditions and events that it may encounter? 

Clearly the answer is negative, and there are many reasons: 
(1)~The system will encounter new objects and systems that might be invented and created only after the system at hand is built and deployed, and the users or owners of the deployed system may not apply relevant updates, if any are available at all. (2)~For reasons of cost and time, developers sometimes choose to not consider certain less likely events, relying on other mitigating factors (like natural help from nearby humans) to handle them. (3)~Successful autonomous systems will be distributed world-wide, and at least some of them are likely to be deployed in environments with which the developers were not familiar and did not consider.  (4)~Autonomous systems will often operate in rich environments characterized by many variables and actors (both humans and systems); even the most sophisticated testing and verification tools we have cannot analyze the exponentially-many combinations of  actions and variable values, both intended and unintended. (5)~Any system is likely to be the target of malicious attacks --- physical or cyber in nature --- by criminals, hackers and foreign bodies; such attacks are likely to seek vulnerabilities in the systems, which by definition are those aspects that the developers did not prepare for and were not expecting. 

Thus, we may ask: is it realistic to expect autonomous systems of the future to be able to handle such unexpected events? After all, they were unexpected! Here, we claim, the answer is positive. This makes some sense, since humans do it all the time, sometimes successfully and sometimes less so (malicious attacks are a special case). Humans are even required to react in certain ways: when an unexpected event occurs a human's reaction is often brought up in a court of law, because, say, someone was hurt, and the judge  will consider what a reasonable person would have done in such a case. 

We believe that systems, like humans, are also required to react properly upon their first encounter with an unexpected situation.
Once next-generation autonomous systems are deployed, the world will not be able to restrict or constrain  innovation, nor  stop the occurrence of unusual combinations of events.  Such systems will have to be updated in an ongoing manner with instructions for dealing correctly with these new situations. The risks involved may well outweigh what society will be able to handle with appropriate insurance, by imposing regulations on people's behavior, or by simply accepting  certain pains or costs in return for the benefits a new system.   

Thus, while the developers must always  study  the system and its target environments extensively in preparation for the future,
there remains the question of how to equip a system to deal with its unexpected encounters. In~\cite{harelMarronSifakis2019autonomics}, Harel, Marron and Sifakis  argue that a new foundation is needed for  engineering next generation autonomous systems, which they term \emph{Autonomics}. The problem of dealing with unpredicted---or even unpredictable---events and conditions, is a central part of the problem.

In Sections~\ref{behaviors},\ref{knowledge},
\ref{social} and \ref{ExploreTheExpected}, we propose several engineering practices that can help the successful handling of  always-impending occurrences of unexpected events and conditions. These are organized in groups: reactive high level behaviors, knowledge that the system must have at its disposal, and considerations emanating from the system's being part of a larger ecosystem of humans and machines.
This preliminary list was compiled from discussions with scientists and  professionals, and reviewed and augmented following an educational workshop for students at the Weizmann Institute of Science. We expect that the list will be extended as part of the systematic construction of the Autonomics engineering foundation.

 Then, in Section~\ref{towardScience}, we propose that the topic of dealing with the unexpected deserves a broader and deeper theoretical underpinning in order to become part of a practical engineering foundation, and we review related research that can support such an endeavor.

\section{Reactive and Proactive Behaviors and Skills}\label{behaviors}

\subsection{Specifying High-level Behavioral Reaction}\label{highLevel}

Throughout life we observe many behavioral patterns. For example, much has been written about Fight or Flight in nature as generic solutions for handling the environment with limited capabilities for sensing,  physical acting, and cognitive processing. In the world of science fiction, Asimov suggested generic laws to govern all robotic behavior, related to not injuring others, obedience, and self-preservation.  Certain social norms and etiquette, as well as business rules, form another facet of practical, yet high level,  programming of behavior. We propose that one can even today build into the system concrete behavioral rules that use abstract concepts, like ``When sensing danger, get away'', ``when under attack,  find shelter'', ``when you cannot understand what is happening, slow down''. Separate definitions can classify  sensory data indicators as `danger' `attack' or  `not understood', and assign various actuator capabilities for implementing the reactions of `getting away', `finding shelter', or `slowing down'. For example, for an autonomous vehicle (AV) on a highway, slowing down may mean  reducing speed, whereas for a robot or a delivery vehicle  on a factory floor or plant yard (FFAV)  `slowing down' may be implemented as an immediate stop. For getting away from danger on a city street, an AV could make a U-turn, whereas an FFAV may simply switch  into reverse gear. 
Rules like these were probably in the backs of the minds of first-line drivers in Figure~\ref{fig:tsunami}, who  saw the tsunami wave and immediately turned back. 

As part of such concrete, yet high-level behaviors, the system may have not only  goals and constraints, which are common concepts in system design, but also \textit{responsibilities}. For example, an autonomous taxi has goals (e.g., to transport passengers between  two points), and requirements (e.g., to comply with traffic and safety rules). However, if we also program into the system the more general responsibility for  passenger safety,  additional behaviors may emerge when  unexpected events occur. We might want such a new behavior to mimic what happens when a person drives a friend in a foreign city and some major failure or traffic jam occurs; namely, the driver is expected to help the passenger find an alternative means of transportation, and not merely abort the goal and abandon them.

\subsection{Probing}
When encountering an unfamiliar situation, a system can actively explore what it means. For example, assume that an FFAV's narrow route in the plant yard is blocked by a large unknown object. The FFAV should be able to figure out if the object is just an empty cardboard box or plastic bag  blown over by the wind. Standard cameras and  basic actuators, like a robotic arm, combined with internet image lookup should be enough for detecting this fact, maybe accompanied by slight  prodding of the object by the FFAV body. The robot can then decide whether to push the box or bag aside, pass around it, or call for help.
An important design implication here is that for probing certain unexpected objects and situations the system may need to be equipped in advance with additional sensors and actuators, or have dynamic access to other such facilities (like security cameras or other robots), beyond the minimum required for completing its missions in typical environments. 

\subsection{[Self-]Reflection}
The system should be able to ``look'' at itself and recognize its own state and history,  using this information in its decision making.   
Unlike a poor fly repeatedly attempting to escape through a glass window, an autonomous system should be able to notice that it has been in the current state before, or that a certain action did not yield the desired results, and apply appropriate actions. For example, from the movie ~\cite{MIT2015DARPAChallenge} it appears that the MIT robot in the DARPA competition of 2015 begins to shake, loses its balance and falls when it tries to step out of a still-moving vehicle. In addition to standard events indicating arrival at the destination and that it is time to step out of the car, it is possible that having the robot sense that it is still moving forward, or sensing that it is beginning to shake, could be used to prevent the actual damaging fall. 

The reflection can extend far into the past. Given a current situation, the system can examine historical event logs to determine which actions yielded desired results in similar situations. 

To enrich state awareness, designers can pre-specify  finer levels or order of desirability among states. The system can then apply these in its planning decisions; see e.g., the discussion of soft goals in~\cite{SofferWand2005SoftGoals}. 

Reflection applies also to broader context. For example, An autonomous system may need to be aware that it should be quiet, say, late at night, or when someone is having an important conversation,  and consequently  change its action plans. Or, an AV trained to drive on the right, may realize that it is in country where driving is on the left, and determine that it should stop completely and/or turn over control to someone else.  

An important form of reflection is understanding causality relationship between system actions and expected and unexpected results. This can help in handling such unexpected situations - which may stem either from (errors in) system decisions or from environmental changes.

\subsection{Physical and Logical Look-Ahead}\label{lookahead}
The farther the system can look ahead in space and in time, the less unexpected the future will turn out to be. A high-mounted 360-degree camera can provide a better view for an AV than the standard driver view. Access to plant security cameras can provide an FFAV with an up-to-date real-time map of all stationary and moving objects relevant to its path and mission. Run-time simulation can provide predictions, e.g., whether the  stationary car that just turned on its lights will start moving, or whether the system's battery has enough charge to complete the current mission. 

\subsection{Preparing and Pursuing Alternative Solutions}
Autonomous systems should be designed with several alternative solutions for completing their tasks, in addition to component redundancy and extra capacity that can compensate for failure or performance constraints. 

Relating to the example of a taxi system and a foreign passenger (in Section~\ref{highLevel}),  
the system should not only be able to find alternative routes when it encounters traffic jams and blocked roads. Should the system realize that it cannot complete its mission, it should be able to automatically transfer the passenger into another mode of transportation---such as taking them to a train station or calling another taxi on their behalf.

\section{Knowledge and Skill Acquisition}\label{knowledge}

\subsection{Knowledge of the system's own capabilities} \label{knowCapabilities}
The availability of the system's sensors and actuators should be ``known'' generally, and be accessible to multiple components rather then be  invoked only by mission-specific code. In this way, scenarios for, e.g., providing camera views to other programmed functions or to a human assistant, or operating the motors upon demand to handle unexpected conditions, could emerge from automated planning when facing new conditions. Agent design concepts such as Belief-Desire-Intention (BDI)~\cite{georgeffTambeWoolridge1998beliefDesireIntentionBDI} can help implement such approaches. Such knowledge can also be useful when the autonomous system is called upon to help other systems, carrying out functions beyond its original ones. 

Knowing the system's own capabilities involves also confirming their existence and proper operations. If a component has already failed, it is better to proactively discover it in advance, rather than wait until the capability is needed.  

\subsection{Access to General World Knowledge}
For carrying out many of the unexpected functions, the system must have common knowledge about the world, and  ``laws of nature'' in the environment in which it operates: from a physics engine that understands gravity, speed, friction and impact , to predicting object behavior (e.g., stationary boxes on the floor don't start moving on their own; people, on the other hand do; the speed of a human walking is at most X, and the speed of an electric scooter can reach Y in a few seconds). This knowledge can be made local to the autonomous system, or made accessible from other systems or servers.

\subsection{Automated Run-time Knowledge Acquisition}
Systems can obtain valuable information in real time. One can only wonder if any of the drivers turning back in Figure~\ref{fig:tsunami} did so not because of what they saw, but because they just heard a tsunami warning siren or announcement. Collecting weather and road condition information, AVs can thus `predict' that their planned route will be impassable due to a snowstorm forecast, and an FFAV checking the locations of humans in a plant may discover that there is no person to receive the current delivery, and perhaps reschedule the entire trip. This aspect is, of course, closely related to the look-ahead discussed in Section~\ref{lookahead}. 

\subsection{Learning and Adaptivity}

Having systems learn from their own successes and failures and other aspects of experience is a key theme in modern engineering and is crucial also to handling unexpected events. We propose that learning and adaptivity can be applied in a broader way. For example, if an AV or a system of AVs also learns and shares knowledge about the behavior of the objects  they see in their environment, such as strange agricultural machinery in a field, it is possible that an AV would handle better the very first situation in which they encounter such an object. 

\section{Viewing the System as a Social Entity}\label{social}

Humans often deal with 
unexpected circumstances by relying on various forms of interactions with others.
In this section, we discuss design considerations and proactive work of the system, which mimic such social reliance and collaboration, potentially reducing unexpected circumstances and facilitating handling them when they do occur. 

\subsection{Delineating the System's Scope of Responsibility}
``No Man is an Island'', says John Donne. The same holds for systems, especially autonomous ones.
Relating to the earlier discussion on responsibilities, it is important to articulate roles and responsibilities. Who is responsible for getting an executive from an airport in a foreign city to an important meeting in town, safely and on time? The autonomous taxi itself? The taxi company? The executive himself or herself? 
The company for which the executive works and has arranged  the trip? 
Moreover, if there is a problem with the AV and alternatives are sought, whose responsibility is it to provide another vehicle, to arrange for another mode of transport, to decide to send someone else to the meeting, or to arrange for a teleconference instead? Determining these roles in advance can be very useful in handling the unexpected. 

Given some `responsibility space' in a problem domain (see discussion of ontologies in Section~\ref{towardScience}) one may even consider formal-verification approaches to help close any gaps in assignment of responsibilities and roles. 
 
\subsection{Mimicking Others}
Some of drivers in Figure~\ref{fig:tsunami} probably did not see the tsunami wave themselves and did not hear about it on the radio, but reacted to the other drivers' sudden reversal and flight. 
Systems should be able to apply such mimicking too, when appropriate. For example,  
if an AV proceeding along a two-lane road notices that it own lane is completely empty but that the other one consists of a long line of queued cars, it should probably mimic the queued cars and ease into their lane, "realizing" that the empty lane is blocked and should not be used.   

\subsection{Asking for Help and Support}

Prompting a human for aid at important decision points or in carrying out tasks that the system cannot do itself, is a standard and common practice in system design, so it is only natural that autonomous systems facing unexpected circumstances should do the same. However, this latter context requires special attention and perhaps different design. Examples of question that have to be addressed are: Who should be contacted when such a situation requires outside help, a human or another machine? Should the contacted agent (human or machine) be in a remote central control function or physically close to the system? What skills and capabilities must the helping agent have? What information should the autonomous system communicate to the selected helper? What interfaces are needed in the autonomous system to enable such help?  
(Having the system apply its own capabilities to help others is discussed in Sections~\ref{knowCapabilities} and~\ref{sharingInformation}.)

\subsection{Enabling Passive Acceptance of Help}
An autonomous system may not always be aware that it is in need of help, but should nevertheless facilitate it. A passerby should be able to communicate to an AV that there is some alarming danger ahead; a control center should be able to take over and override systems decisions---either forcing certain behaviors or a complete standstill; or if the system is stuck in a place with no outside communication it should be able to communicate (through computer display or local data communications, or just a fixed printed sticker) enough information about its identity (and possibly its current state), so that other agents can act at their own initiative, or transmit the information onward to the owners of the system from another location. 

\subsection{Communicating Overall Plans and Present Intentions}

Turn signals and brake lights are simple facilities that indicate a driver's intention and thereby help prevent  accidents and reduce  stress.
Communication, coordination and negotiations are already a common design aspect in multi-agent systems~\cite{kaminka2013curingRobotAutism, georgeffTambeWoolridge1998beliefDesireIntentionBDI}. 
Autonomous systems can communicate their intentions much more broadly, with great effect. For example, an AV can communicate its planned route, thus helping navigation systems predict road congestion; it can also indicate from far away when it plans to stop at a crosswalk, allowing pedestrians to move forward sooner. An industrial robot can communicate its intended next steps, allowing humans to work more safely closer to it.  

\subsection{Recording and Sharing Static and Dynamic Knowledge}\label{sharingInformation}
Trust in the system, and/or in the organization controlling it, can be dramatically enhanced by providing information that contributes to the mitigation and handling of unexpected conditions. 
One aspect is sharing static information about how the system is built and how it is programmed to behave. 
Another is collecting and sharing dynamic information both about the system's own behavior and about its environment. The latter can be passive --- as in sharing security camera recordings --- or  proactive, reporting current conditions, changes relative to known maps, hazards, or even reporting humans and other devices that it has encountered and who may need help. 

\subsection{Negotiations}

Often dealing with the unexpected is difficult because of constraints and specific instructions imposed at design time. 
However, in real life many constraints are bendable. Human drivers often communicate with others in order to resolve issues like being in the wrong lane; people standing in a queue can negotiate to allow one who is in a hurry to proceed; and many people in a room can find ways to have everyone feel comfortable with the temperature, combining opening or closing of windows, settings of the A/C, wearing jackets, and/or rearranging the seating.
When encountering unexpected events and conditions, an autonomous system should be able to participate in such negotiations with the entities in its environment, by communicating its desires, what it can and cannot change, understanding the (changing) needs of others, and adjusting its goals and plans.
Here again, BDI concepts~\cite{georgeffTambeWoolridge1998beliefDesireIntentionBDI} may prove instrumental. 

\section{Proactive Exploration of the Expected}\label{ExploreTheExpected}

Complementing the sections on how to deal with the inevitable occurrence of unexpected circumstances, we briefly review techniques which one can use in exploring the problem domain in anticipation of possible situations, towards making explicit, ordinary preparations: (1)~straightforward studies of the system and its environment using documents, existing systems,    brainstorming with colleagues and domain experts, etc.; (2)~considering the system's operation in related but more challenging environments, e.g., for an AV, instead of an intended modern quiet suburban neighborhood, consider it operating in a busy narrow alley of a medieval town packed with tourists; or, for an FFAV, instead of the originally targeted hospital environment, consider it operating on a factory floor or plant yard; then, reapply relevant new aspects of the more demanding conditions back to the original environment;  (3)~challenging developers and colleagues to create failure situations for the system; (4)~studying famous failures in similar systems and environments; (5)~building models of the environment and the system, running simulations, and then examining simulation runs for emergent properties that can suggest vulnerabilities; (6)~noting in the above rather standard simulations aspects that appear too synthetic or oversimplified and `digging deeper' in those; e.g., if an AV simulation exhibits a very low variability in pedestrian behavior, one can think of what the system should do if it encounters a performing street juggler, or a drunk brawl outside a bar, and do this analysis regardless of whether the current simulator can recreate these conditions or not; (6)~mentally simulating the drafting of a classical legal 
contract with regards to the system's behavior, focusing on using certain terms in advance which can cover broad situations when unexpected situations occur. 

\section{Toward a ``Science of the Unexpected''}\label{towardScience}

We believe that preparing systems to deal with the unexpected should be one of the topics covered by the Autonomics engineering foundation discussed in~\cite{harelMarronSifakis2019autonomics}. Going beyond the specific ideas listed in the preceding sections, the foundation should cover the topic accompanied by systematic engineering methodologies for all phases of development, by reusable specification libraries, domain-specific ontologies, simulation, testing and verification tools, and more. 

 In addition, we believe that there will be a need for a scientifically rigorous apparatus for the domain. This should facilitate detailed and broad study of the unexpected, ensure fresh examination and evolution of the solutions, as systems and the real world evolve, and enable critical assessment of the quality (e.g., completeness, non-ambiguity) of all associated practices and artifacts. 
 
To evolve such theoretical instrumentation, various perspective approaches may be adopted. Below we briefly discuss two, and acknowledge that there is still much more to be investigated.

One possible perspective is to incorporate concepts from  \textit{digital twin} approaches~\cite{saddik2018}, or more fundamentally, to  use a \textit{representational model} as part of the \textit{deep-structure} view of an Information System (IS), as proposed by Wand and Weber~\cite{wandWeber1995}. According to this view, an IS is seen as a digital means for representing some domain or some part of the real-world. For example, a camera complemented by automatic object detection algorithm and data persistence should faithfully represent the elements anticipated to exist in the real-world environment in which the system is deployed.  

This view has also given rise to the adoption of various ontological theories, such as the one by  Bunge~\cite{bunge1978}, for determining the capacity of any IS to faithfully represent its real-world environment. This way, an ontological taxonomy can determine, for example, the validity of any object instantiated in a system, i.e., whether it has  meaning or semantics within the domain, regardless of its concrete classification in the IS. Instead, each such object can belong to a corresponding ontological concept such as a \texttt{thing} or an \texttt{event} in the real world. This approach can assist in formal definitions of the unexpected, such as this: 

\begin{definition}
We say that an object, an event-occurrence or a system state which is associated with a particular concept in a given ontology is \textit{unexpected} in the context of a given system, if either (a) the ontological concept (i.e., class) does not have a representation in the system, or (b) the system cannot effectively associate the object, state, or event-occurrence  with the (existing) system representation of the ontological concept, dynamically, at run time.
\end{definition}

According to the ontological view and the above definition, we can now also conclude that any IS, being an artifact, is always susceptible to \textit{ontological deficit}, where it may create or encounter objects that lack a corresponding interpretation in the system. 
Recognizing this deficit in real time and perhaps transferring the object for handling by humans or by another system can be part of a design for dealing with the unexpected.

Different systems may also differ in the way they implement the classification mechanism. For example, the work by Parsons and Wand~\cite{parsonsWand2000emancipatingInstancesFromTyrannyOfClasses} has criticized the assumption of \textit{inherent classification}, under which an object can only be recorded if it is a member of a certain class. The same implementation also mandates the specification of each class via the collection of its objects (i.e., \textit{classification-by-containment}). An example of a mechanism that conforms to this approach is the traditional \textit{relational database} model and the underlying operations thereof~\cite{codd1970}. 
In object oriented programming, multiple inheritance and interfaces address some of these issues, but the basic classification remains. 

A system that does not conform to the assumption of inherent classification enables the recording of objects independent of their class membership(s). Furthermore, in \cite{parsonsWand2000emancipatingInstancesFromTyrannyOfClasses}, the authors suggest one such alternative, in which a \textit{classification-by-property} approach is employed. This approach entails articulation of objects and of classes independently of each other, and specifying classes and objects via their properties. For example the class(\textit{dangerous-species}) may be defined as any living entity that possesses any one of the properties \{eye-size = big, teeth-size = big, crawling-ability = true\}. Similarly the object(\texttt{snake1}) may be defined as an entity that possesses the properties \{location = \textit{location1}, crawling-ability = true\}. 

Decoupling specifications of classes and objects is essential for dealing with the existence of the unknown, that is, objects in the domain that have no meaning in the system. The importance of a specification as given in the above example is not only in allowing the ad-hoc determination of \textit{snake1} as being an instance of the class \textit{dangerous-species}. It is also in allowing an independent representation of an object like  \textit{snake1}, regardless of whether or not we know its class. An unexpected object produced in or encountered by a system that conforms to the assumption of inherent classification is at risk of being either incorrectly classified or altogether dropped. Both may lead to undesired outcomes.

Another perspective is a socio-technical one, where a human encounters unexpected events or conditions and a system augments the person's capabilities in dealing with the situation~\cite{orlikowski2001}, while mitigating biases. 
For example, a security officer in a large store may be monitoring the store via multiple screens. According to the \textit{representativeness heuristic} introduced by ~\cite{tverskyKahneman1974}, the  officer, as a human being, may  focus on certain kinds of people, based, say, on appearance, as opposed to focusing on actual behavior.

Finally, research in neurology, psychology, philosophy, discusses at length many issues of reconciling reality, perception and expectations (see, e.g.,~\cite{pressKokYon2019perceptualPredictionParadox} and references therein). Advancing insights about the human mind may well feed into our engineering goals.  

\begin{acks}
This work has been supported in part by a grant to David Harel  from Israel Science Foundation, the William Sussman Professorial Chair of Mathematics, and the Estate of Emile Mimran. 
The authors thank Baruch Feldman, Yair Finkelstein, Ayelet Marron, Orley Marron, Anthony Morena,  Fabio Patrizi, and Meir Shani for valuable discussions and suggestions.
\end{acks}

\bibliographystyle{ACM-Reference-Format}
\bibliography{ExpectingTheUnexpectedBib}

\end{document}